# Application Independent Energy Efficient Data Aggregation in Wireless Sensor Networks

Deepali Virmani, Tanuj Singhal, Ghanshyam, Khyati Ahlawat and Noble

Guru Gobind Singh Indraprastha University

Delhi, India

deepalivirmani@gmail.com, khyatiahlawat@yahoo.co.in

**Abstract:** Wireless Sensor networks are dense networks of small, low-cost sensors, which collect and disseminate environmental data and thus facilitate monitoring and controlling of physical environment from remote locations with better accuracy. The major challenge is to achieve energy efficiency during the communication among the nodes. This paper aims at proposing a solution to schedule the node's activities to reduce the energy consumption. We propose the construction of a decentralized lifetime maximizing tree within clusters. We aim at minimizing the distance of transmission with minimization of energy consumption. The sensor network is distributed into clusters based on the close proximity of the nodes. Data transfer among the nodes is done with a hybrid technique of both TDMA/ FDMA which leads to efficient utilization of bandwidth and maximizing throughput.

**I Introduction**

The current technological advancement has already come to terms with immense potential of Wireless Sensor Network, which consists of tiny sensor nodes scattered in a region communicating with each other over well defined protocols and transferring information of temperature, humidity etc among each other. To exploit the potential of WSNs various studies have focused on Data aggregation approach which requires data to be collected and processed at a single node prior to its transfer to the parent node.

Realization of this data aggregation approach has been a major concern owing to the limited battery life of the sensors which limits the lifetime for which a sensor node remains active. The limited battery leads to disruption in the connections from the network. These disruptions suggest that the design must incorporate topological changes. Thus, idea is not to allow the direct transmission of to interested users upon event detection instead aggregating them to remove redundancy. The application domain of WSN is still expanding therefore it is important to support the data aggregation scheme from multiple nodes for simultaneous and fast processing of the data.

In this paper, we focus on the energy efficiency of data aggregation, and propose a new and effective data aggregation scheme based on clustering. Our scheme consists of three parts, namely, the clustering of nodes by using the Expectation-maximization (EM) [9] algorithm, construction of life maximizing tree structure using MST_PRIMS [8] and aggregating the data collected from the WSN nodes by applying a cluster scheduling approach to transfer it , which uses TDMA/FDMA[1] mechanism.

The rest of the paper is organized as follows. Section II describes some related works. Section III elucidates our approach. Section IV has the proposed algorithm. Finally, concluding remarks and future work are provided in Section V.

## II Related Works

In recent literature many studies have achieved data aggregation using several approaches namely mobile sink , LEACH(Low-Energy Adaptive Clustering Hierarchy), Directed Diffusion [9].All these schemes have tried to prolong the network lifetime and reduce the energy consumption The mobile sink scheme increases the network lifetime to four times as compared to the network in which sink is static but it suffers from serious shortcomings , it leads to an increased physical delay owing to the slow physical mobility of the sink then the wireless communication. It exhausts the battery life unnecessarily.

LEACH is a self-organizing, adaptive clustering protocol. To have minimum energy consumption, nodes in LEACH are grouped into a number of clusters based on their battery usage. Each cluster has a cluster head, which communicates with every node of that cluster. The sink aggregates data, transmitted by cluster heads, from other nodes. Since a cluster head loses energy due to repeated transmissions, the cluster head is re-selected based on the residual energy, as a consequence it prolongs the network lifetime. Directed Diffusion involves two types of messages, namely, the "Interest" message and the actual data messages. To aggregate data by using Directed Diffusion, the sink node broadcasts an "Interest" message that consists of a time-to-live value, and also the addresses of the source and destination nodes. The destination node on receiving the request transmits appropriate data message to the source having the sensed data. If the downstream nodes cannot be reached by the "interest" message from the current source then the current destination becomes the source node by changes its address, reduces the time –to-live value and rebroadcasts the "Interest "message.

## III OUR APPROACH

We have discussed before that our main focus has been on distance minimization between the nodes, energy efficiency and efficient utilization of bandwidth. Considering the problem sensor network is divided into clusters using an EM [9] algorithm based on the close proximity of the nodes and a life maximizing [8] tree is constructed within the cluster choosing a parent closest to the sink node to serve as sub-sink. Once the sub sink is chosen scheduling of cluster is done using FDMA approach assigning a range of frequencies from the available one to the sub-sink. The frequency ranges can be re-allotted to new clusters from the free pool or assigning a half frequency to a sub-sink from a low data rate transferring cluster. The sink will broadcast a topology packet containing information of the network as which source nodes are attached to which sub-sink node as per their location [4]. By making use of the hybrid TDMA/FDMA channel access technique [1], the sink node broadcasts a schedule packet informing others about their time slots as well as their channel frequencies for exchanging messages. But, our idea lies in single sleep awake concept in which the source nodes wake up only once to listen and to transmit and rest of the time, they will remain in sleep state. We incorporate a concept of LPL (low power listening),the nodes are in LPL [2] state all the time to gauge topology changes and if there is a topology packet coming their way they wake up and make necessary changes if a new neighbor has come up and further retransmit the same to their neighbors. The tree construction follows a life maximizing tree MST_PRIMS[8] which builds a tree using a key 'K' which is the ratio of residual energy of the nodes to the distance between the nodes. .To overcome the problem in the HyMAC[1] and energy efficiency wake up scheduling for data collection [5] and scheduling [7] of increase in the number of nodes leading to increased slots in FDMA to decrease the interference respectively. We will assign specific frequency slots based on attributes [6] of the sensor nodes, with fixed interference ranges so that they can send their data in scheduled time in slotted frequency. Once the sender finishes sending, same frequency can be assigned to some other source accounting for the increase slots needed and also minimizing interferences.

To preserve the functional lifetime of all sources and efficient utilization of the energy of the source nodes the proposed MST_PRIMS construction algorithm arranges all nodes in a way that each parent will have the maximal-available energy resources to receive data from all of its children. Such arrangement extends the time to refresh the tree and lowers the amount of data lost due to a broken tree link before the tree reconstructions.

The MST_PRIMS [8] algorithm can be further improved by considering distance also as a factor. We would like to include distance between the sensor nodes. Transmission distance has a major impact on the working of sensor network because the required power of wireless transmission is proportional to the square of the transmission distance. We follow an approach of clustering of nodes based on EM [9] algorithm. The EM algorithm includes minimizing the sum of the squares of the distances between nodes and cluster centroids. Therefore, we use the EM [9] algorithm to group the WSN nodes into K clusters on the basis of distance. We apply the concept of EM [9] algorithm initially and then use a new form of life maximizing tree, MST_PRIMS [8] algorithm accordingly. The cluster formed using the EMD algorithm go through our proposed algorithm called MST_PRIMS on energy and distance (MST_PRIMS), which creates trees within the clusters already created .The choice of the tree is based on the minimum distance of the sub-sink from the sink.

The tree that we get after application of both the algorithms is efficient in terms of distance as well as energy, now to improve the bandwidth utilization we apply a method of FDMA_SINK() that allots range of frequencies to the sub-sink for data transfer and also checks for the efficient utilization of bandwidth by assigning half frequency range from a low data transfer cluster to a new one.

Finally, after frequency allotment we implement HyMac [1] algorithm which provides fixed time slots to nodes to transmit sensed data, the sub-sink remains in a LPL state and listens for receiving data from the children, the child nodes awake once and start the synchronous data transfer to the sub-sink which further sends sensed data to the sink in the same assigned time slot.

## IV PROPOSED ALGORITHM

### A: EM Algorithm

The following algorithm that we use divides the network into K clusters. It has been renamed from EM [9] to EMD (Expectation Maximization on Distance) Algorithm.

**K**: The number of clusters

**Π$k$**: The mixing coefficients of the $k^{th}$ cluster

**µ$_k$**: The 2-dimensional vector indicates the mean of the $k^{th}$ cluster

**Σ$_k$**: The $2 \times 2$ covariance matrix of the $k^{th}$ cluster

---

**Algorithm 1: EMD Algorithm**

The mobile sink node groups all nodes into K clusters by using the EMD algorithm in the following manner.

1: Initialize µ, Σ, π and the convergence criterion $\theta_{EM}$, and evaluate the initial value of P:

$$P = \sum_{n=1}^{N} ln \left\{ \sum_{k=1}^{K} \pi_k N(x_n | \mu_k, \Sigma_k) \right\}$$

Where N is the number of nodes.

---

The above presented algorithm groups the sensor nodes into clusters based on their distance and hence ensure that the nodes in a cluster have a close proximity which will lead to minimization of data transfer delays. The algorithm can be iteratively used to account for any new nodes coming up in the network.

### B: Life maximizing tree using MST_PRIMS algorithm

Description: Let G = (V, E) be a weighted, connected graph. Let T be the edge set that is grown in Prim's algorithm. The proof is by mathematical induction on the number of edges in T and using the MST Lemma. The empty set Ø is promising since a connected, weighted graph always has at least one MST.

Assume that T is promising just before the algorithm adds a new edge e = (u,v). Let U be the set of nodes grown in Prim's algorithm. Then all three conditions in the MST Lemma are satisfied and therefore T U e is also promising.

When the algorithm stops, U includes all vertices of the graph and hence T is a spanning tree. Since T is also promising, it will be a MST.

We consider a Key, 'K' to build the tree which is the ratio of residual energy of the node to distance between the nodes.

## C: Applying FDMA_SINK on the MST_PRIMS Tree

**Scheduling using FDMA System Model**:

All the frequencies available in the band are divided in frequency ranges(R) based on the number of clusters(K) ,Initially, all sub-sinks send a synchronous message to sink requesting allotment of the frequencies ,sink then checks for the availability of the frequency ranges. If it is available it is assigned. Otherwise sink may withdraw a frequency from some other cluster if its data transmission rate is low and assign that to some cluster. We propose an algorithm FDMA_SINK() with parameters (K ,number of clusters,Fi ,frequency band, i ,Cluster),SSi is subsink of ith cluster.

## D: Cluster Scheduling Using TDMA/FDMA

Scheduling algorithm is applied on the MST_PRIMS tree having the base node as its root. As each node Ni is traversed by MST_PRIMS, it is assigned a default time slot and a frequency using FDMA_SINK discussed before. Then the possibility of having an interference with any of its same-height previously-visited one-hop AND two-hop neighbours is checked. If a conflicting neighbour Nj is found for Ni, the algorithm checks whether Ni and Nj are siblings. If so, Ni will be assigned a different time slot than that of Nj. If they are not siblings then Ni will be assigned a different frequency than that of Nj, allowing both Ni and Nj to send messages to their parents at the same time slot but in different channels. When MST_PRIMS is about to start a new level (height) of nodes the default time slot number will be increased by one. Once all nodes are processed according to the above heuristic, the entire time slot assignments will be inverted such that the slot number assigned to every node is smaller than that of its parent. This inversion is done as following:

$tnew = tmax − tcurrent + 1$

Where tnew is the new inverted assigned slot, tcurrent is the current slot number assigned to the node and tmax is the total number of assigned slots. Note that such an assignment allows the data packets to be aggregated and propagated in a cascading manner to the base station in a single TDMA cycle. The complete steps of the overall process are presented in algorithm 4.

---

**FINAL ALGORITHM:**

**Require:** Set of sensor nodes

Step 1: Apply EMD() /* Creates clusters */

Step 2: Creating Life maximizing tree using MST_PRIMS

Step 3: Scheduling Using HYMACED which calls FDMA_SINK for scheduling based on time slots and frequency range.

---

## V. Conclusions and future work

**Conclusions**

Clustering on the basis of distance ensures close proximity of the nodes and thus leads to reduction in data transfer delays. Energy conservation by waking nodes once instead of twice leads [5] to further reduction in data transfer delays, thus utilizing the energy available effectively.

Efficient utilization of bandwidth achieved through allotment of frequencies from the free pool or withdrawing half frequency from cluster with low data transfer and assigning it some other.

**Future Work**

The algorithm needs to be implemented and results needs to be verified in a Simulator.